# Improving Reproducibility and Performance of Radiomics in Low Dose CT using Cycle GANs

**Running Title:** Improving Radiomics Using Cycle GANs

Junhua Chen [a,1], MS; Leonard Wee [1], PhD; Andre Dekker, PhD[1]; Inigo Bermejo, PhD[1]

[1]. Department of Radiation Oncology (MAASTRO), GROW School for Oncology and Developmental Biology, Maastricht University Medical Centre+, Maastricht, 6229 ET, Netherlands

[a] Corresponding author, Tel.: +31 0684 6149 49 E-mail address: j.chen@maastrichtuniversity.nl

**Address of Corresponding author:** Department of Radiation Oncology (MAASTRO), GROW School for Oncology and Developmental Biology, Maastricht University Medical Centre+, Maastricht, 6229 ET, Netherlands

**Abstract:**

**Background:** As a means to extract biomarkers from medical imaging, radiomics has attracted increased attention from researchers. However, reproducibility and performance of radiomics in low dose CT scans are still poor, mostly due to noise. Deep learning generative models can be used to denoise these images and in turn improve radiomics' reproducibility and performance. However, most generative models are trained on paired data, which can be difficult or impossible to collect.

**Purpose:** In this article, we investigate the possibility of denoising low dose CTs using cycle generative adversarial networks (GANs) to improve radiomics reproducibility and performance based on unpaired datasets.

**Methods and Materials:** Two cycle GANs were trained: 1) from paired data, by simulating low dose CTs (i.e., introducing noise) from high dose CTs; and 2) from unpaired real low dose CTs. To accelerate convergence, during GAN training, a slice-paired training strategy was introduced. The trained GANs were applied to three scenarios: 1) improving radiomics reproducibility in simulated low dose CT images and 2) same-day repeat low dose CTs (RIDER dataset) and 3) improving radiomics performance in survival prediction. Cycle GAN results were compared with a conditional GAN (CGAN) and an encoder-decoder network (EDN) trained on simulated paired data.

**Results:** The cycle GAN trained on simulated data improved concordance correlation coefficients (CCC) of radiomic features from 0.87 [95%CI, (0.833,0.901)] to 0.93 [95%CI, (0.916,0.949)] on simulated noise CT and from 0.89 [95%CI, (0.881,0.914)] to 0.92 [95%CI, (0.908,0.937)] on RIDER dataset, as well improving the area under the receiver operating characteristic curve (AUC) of survival prediction from 0.52 [95%CI, (0.511,0.538)] to 0.59 [95%CI, (0.578,0.602)]. The cycle GAN trained on real data increased the CCCs of features in RIDER to 0.95 [95%CI, (0.933,0.961)] and the AUC of survival prediction to 0.58 [95%CI, (0.576,0.596)].

**Conclusion:** The results show that cycle GANs trained on both simulated and real data can improve radiomics' reproducibility and performance in low dose CT and achieve similar results compared to CGANs and EDNs.

## 1. Introduction

Biomarkers from medical imaging can provide a macroscopic view of the tissue of interest and can be an effective tool to accurately diagnose disease in precision medicine [1]. Radiomics features [2] have shown value as potential imaging biomarkers in various tumor and neurodegenerative diseases, such as lung cancer [3], head and neck cancer [4], rectal cancer [5], breast cancer [6], Alzheimer disease [7], autism spectrum disorder [8] etc.

However, in Computed Tomography (CT) the repeatability and reproducibility of radiomics has been challenged in multiple published studies [9][10][11][12]. Reproducibility of radiomics can be impacted by various parameters of CT such as radiation dose, slice thicknesses, reconstruction algorithm settings, etc. More specific, only 11.3% (12 of 106) of radiomics features were reported to be robust for all different technical parameters (dose level, reconstructed slice thickness, reconstruction kernel and algorithm) [12]. Of these, intensity Radiomic Features and Texture Radiomic Features are highly sensitive to radiation dose and the associated signal to noise ratio[12]. Therefore, it is likely that radiomics extracted from low dose CT are less accurate than features from high dose CT. In other words, radiomics applied in low dose CT will likely have a low reliability and thus the established radiomics signature or models are likely to have a worse performance compared to high dose CT [13].

In this study, we aim to use denoising [14] to improve the reliability of radiomics in low dose CT. A variety of image denoising methods have been proposed in the past several decades, and these methods can be divided into two classes -- model based denoisers [15][16] and data driven denoisers [17][18]. Multiple published studies [18][19] have demonstrated that data driven denoisers have a better performance compared to model based denoisers and achieve the state-of-art denoising quality if suitable training datasets are available.

Most data driven denoisers are based on deep convolutional neural networks (DCNNs) [20] in which this denoising task is posed as an image-to-image translation problem. The popular architectures for medical image denoising are full convolutional network (FCN) [21], encoder-decoder network (EDN) [22] and generative adversarial networks (GAN) [23] which were described in detail recently reviews [14][24]. An important characteristic of most data driven denoisers is that datasets consisting of paired low-high dose CTs from the same subjects are needed to train the deep neural networks. However, collecting paired low-high dose CT is time-consuming, expensive, and impossible in many cases e.g., in patient studies.

Therefore, it is the aim of this study to establish a CT denoiser based on unpaired datasets to improve radiomics performance. The related literature is divided into two topics -- low dose CT denoising and radiomics normalization. In this section, we review these two topics briefly.

a) Low Dose CT Denoising

As mentioned above, most data-driven denoisers are based on one of three backbones – FCN, encoder-decoder network and GAN – and all of them are used in low dose CT denoising tasks. More specifically, Yang et al. [28] used a 3D residual network as the denoising network architecture with a loss function based on differences between the ground truth residual image and reconstructed residual image. Moreover, pool layers were removed from the network to generate denoised residual images because there is no size or resolution change between input and output. The results show that the network can reduce noise effectively while preserving tissue details. Chen et al [29] adapted an encoder-decoder network as the backbone of their denoiser and two residual shortcuts were added into the network to keep details of the image from encoder to decoder. Models were trained by using simulation data and the trained denoiser achieved a competitive performance in both simulation and clinical cases. Yang et al. [30] took conditional GAN (CGAN) [31] as the backbone where they replaced Jensen–Shannon divergence [32] with Wasserstein distance [33] to measure the differences in the data distribution. Moreover, Yang et al. replaced the mean squared error (MSE) loss function with Perceptual Loss [34] to keep more texture information from low dose CT to high dose CT. They proposed a method to not only reduce the image noise level but also tried to keep the critical information at the same time.

One of the biggest shortcomings of these aforementioned denoisers is that paired low-high dose datasets are needed in denoiser training. However, collecting this kind of datasets is time-consuming and expensive. As an alternative a few simulation paired low-high dose CT datasets are publicly available, such as the dataset from 2016 NIH-AAPM-Mayo Clinic Low Dose CT Grand Challenge (LDGC) [35]. The low dose CT images in this dataset are simulation data with a simulated low radiation dose of 50 mAs. The characteristics of LDGC dataset decrease the value for network training as the generalization of models trained from the LDGC to real low dose CT is questionable because the exposure in real low dose CT datasets will much lower than the simulation data in LDGC,. For example, radiation dose in The Reference Image Database to Evaluate Therapy Response (RIDER) [36] ranged from 7 to 13 mAs.

Therefore, we believe that implementing a denoiser based on unpaired datasets could help to relieve the problem of data collection and make unsupervised CT denoising for quantitative medical image analysis possible. There are a few studies that used this strategy, Kang et al. [37] used cycle GAN as the backbone for multiphase coronary CT angiography correction where they took routine-dose CT from multiphase coronary CT angiography as the target domain data and low-dose CT as the original domain data to build a training dataset. The results show that visual grading and quality evaluation of low-dose CT are improved, however, they did not investigate the effect of Cycle GAN into deeper quantitative metrics such as Radiomics.

However, to the best of our knowledge there are no studies that apply unsupervised CT denoising to improve radiomics reliability and reproducibility in low dose CT.

b) Radiomics Normalization

Berenguer et al. [10] have shown that over half of radiomics features are nonreproducible when images scanned from different scanners even when using the same CT parameters. The results of radiomics signatures or models which based on nonreproducible features are thus unreliable. Li et al. [25] used cycle GAN to normalize CT images from multiple centres and multiple scanners, and then they extracted features from normalized images and established radiomics signatures. They found the average improvement of a classifier based on normalized radiomics features in the area under the receiver operating characteristic curve (AUC) to be 11%.

In previous work, we used EDN and CGAN [13] as testing backbones to denoise low-dose CT. Our training datasets consisted of paired simulated low-dose CT and high-dose CTs. Radiomics features reproducibility from noisy images and denoised images were measured using concordance correlation coefficients (CCC) [39]. The results showed that encoder-decoder network and CGAN can improve CCC of noisy images significantly. Moreover, when we applied our trained denoisers to real low-dose CT images (RIDER dataset), the results showed that this denoiser can improve radiomics reproducibility in realistic low-dose CTs.

In another study [26], we applied the trained denoisers to improve radiomics performance in realistic applications. The results showed that generative models based denoisers can improve the AUC of a lung cancer survival prediction from 0.52 [95%CI, (0.511,0.538)] to 0.58 [95%CI, (0.564,0.596)] and a multiple instance learning based lung cancer diagnostic [40] from 0.84 [95%CI, (0.828,0.856)] to 0.88 [95%CI, (0.866,0.892)].

The major shortcoming of our previous studies is that denoising models were exclusively dependent on paired simulation data which may cause the trained denoiser to not generalize well to real data. In this paper, we took cycle

GAN as basic denoising model to train a denoiser using unpaired low-high dose CT. These low and high dose CT images were collected from different centres and scanners. We evaluated this new denoiser for its ability to improve radiomics reproducibility and performance in realistic applications. Source code, Radiomics features, data for statistical analysis and supplementary materials of this article will be available online at https://gitlab.com/UM-CDS/low-dose-ct-denoising/-/tree/Cycle_GAN_Improve_Radiomics.

## 2. Materials and Methods

In this section, we describe the architecture and technical details of our cycle GAN. Then, we introduce our training strategy to improve the speed of convergence. Next, we describe the design of the experiments and datasets used for training and testing. Finally, we describe the extraction of the radiomics features and the evaluation metrics used.

### 2.1 Cycle GAN

We use cycle-consistent GANs, proposed by Zhu et al. [27]. As shown in Figure 1(a), the cycle GAN consist of two generators and two discriminators. The generator $G_{LH}$ maps from low dose CT domain ($L$) to full dose CT domain ($H$) while $G_{HL}$ maps from $H$ to $L$. The loss function of the cycle GAN consists of two parts -- adversarial loss and cycle consistency loss, represented with $L_{adv}$ and $L_{cyc}$ respectively (and each of them can be broken down into $L_{adv1}, L_{adv2}$ and $L_{cyc1} L_{cyc2}$, one for each generator). The adversarial loss for mapping from low dose to full dose CT is defined as follows:

$$\mathcal{L}_{adv1}(G_{LH}, D_H, L, H) = \mathbb{E}_{h \sim pdata(h)}[log D_H(h)] + \mathbb{E}_{l \sim pdata(l)}[log(1 - D_H(G(l)))] \qquad (1)$$

where $G_{LH}$ is trained to transform low dose CT image $x_l$ to into high dose CT image $x_h$ (denoising), while $D_H$ is trained to discriminate between denoised CT images $G_{LH}(x_l)$ ($x_{LH}$ in Figure 1 (a)) and real high dose CT image $x_H$. During the training, $G$ aims to minimize this loss function against an adversary $D$ that tries to maximize it; therefore, equation (1) can be rewritten as follows:

$$min_G max_D \mathcal{L}_{adv1}(G_{LH}, D_H, L, H) = \mathbb{E}_{h \sim pdata(h)}[log D_H(h)] + \mathbb{E}_{l \sim pdata(l)}[log(1 - D_H(G(l)))] \qquad (2)$$

The definition of adversarial loss for mapping from high dose CT to low dose CT is defined in similar way and we denote it as $min_G max_D \mathcal{L}_{adv2}(G_{HL}, D_L, H, L)$. Moreover, we denote the adversarial loss for the whole network as $\mathcal{L}_{adv}(G, D) = L_{adv1} + L_{adv2}$.

Regarding the cycle consistency loss of our cycle GAN, we replace the mean squared error (MSE) loss function used in the original cycle GAN with a perceptual loss-based loss function. The definition of cycle consistency loss is as follows:

$$\mathcal{L}_{cyc1} = \mathbb{E}(x_l, x_{lh}) \left[ \frac{1}{wed} \left\| VGG\big(G_{HL}(x_{lh})\big) - VGG(x_l) \right\|^2 \right] \tag{3}$$

where $w$, $e$, and $d$ represent width, height, and depth of the feature map, and $VGG(.)$ represents feature maps from a pre-trained VGG-16 at a specific convolutional layer. In our implementation, we select feature maps from $conv2\_1$ to calculate perceptual loss. $\mathcal{L}_{cyc2}$ can be defined in similar way with $G_{LH}$. We denote $\mathcal{L}_{cyc1} + \mathcal{L}_{cyc2}$ as $\mathcal{L}_{cyc}(G)$.

Combining equation (2) and (3), the overall loss function is expressed as:

$$min_G max_D \mathcal{L}_{adv}(G, D) + \lambda \mathcal{L}_{cyc}(G) \tag{4}$$

where $\lambda$ is a parameter to control the trade-off between the adversarial and perceptual loss.

More details about the architecture of generators and discriminators can be found in Figure 1 (b) and (c) respectively.

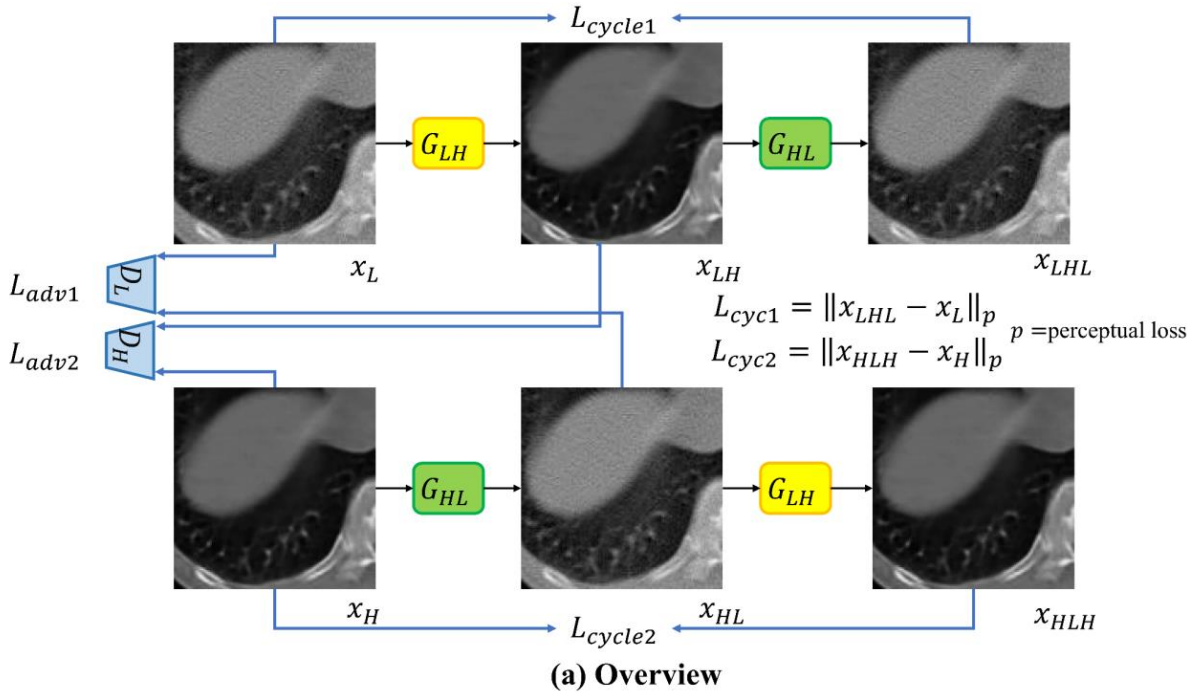

**(a) Overview**

Figure 1. Overview of Network, Architecture of Generator and Discriminator

*2.2 Slice-paired Training Strategy*

Generally speaking, when training a cycle GAN, randomly chosen samples from two domains are fed to the networks. However, as mentioned in the original cycle GAN article [27], the training will be more successful and stable when focusing on pairs of visually similar images.

In the case of CT scans, assuming all scans belong to the same organ (the lung in our case), we can expect that images belonging to the same slice number will be more similar to each other than images from different slices. Hence, the first slice of a low dose CT scan will have higher similarity with the first slice of a high dose CT scan.

Therefore, CT based cycle GAN training should be fed with pairs of the same (randomly chosen) slice rather than images of different slices. This could be seen as weakly supervised learning. We call this strategy as slice-paired training strategy hereafter, the similar training strategy can be found in paper [41].

*2.3 Data Acquisition*

In order to compare results of cycle GANs with our previous work (CGAN and EDN) [13][26], we trained networks on the same data as used in [13][26] and applied the trained models to the same applications on the same datasets. In total, we used five datasets in this study.

The first is based on the NSCLC-Radiomics dataset (hereafter called LUNG 1). We selected only the high dose CT scans, those scanned at 400 milliampere-seconds (mAs) or more (n=157, indices in Supplementary Table 1) and added noise to the sinograms to simulate low dose CTs with two different levels of noise: low-noise CT and high-noise CT. The specific methods used to add noise are described in [13]. We used a subset of these high-noise CTs and their corresponding high dose CTs (40 subjects, 4260 images) to train a cycle GAN and we used the remaining images to assess the reproducibility of radiomics features in the original high dose CT versus those in the denoised images.

The second and third datasets were used to train the cycle GAN with real low dose CT scans. We used low dose CT

scans from the Lung Image Database Consortium dataset (LIDC-IDRI) [44], and high dose CT scans from The Cancer

Genome Atlas Lung Adenocarcinoma (TCGA-LUAD) dataset [45]. We used two inclusion criteria for CTs in both datasets

to increase the visual similarity across the two domains: the use of SIEMENS scanner; table height ranging from 150 to

160 mm. As low dose CTs we included those with a radiation exposure lower than 10 mAs and as high dose CTs those with and exposure higher than 100 mAs (list of indices of selected samples is in Supplementary Tables 2 and 3 respectively).

Examples of selected samples from LIDC-IDRI and TCGA-LUAD are shown in Supplementary Figure 1.

The final two datasets, used for the two radiomics-based applications, are RIDER [36] and NSCLC Radiogenomics

[43]. RIDER is a collection of same day repeat CT scans collected to assess the variability of tumor measurements, which

makes it particularly useful to assess the reproducibility of radiomics across pairs of similar CT scans. We use the trained cycle GAN to denoise the images in RIDER to assess the impact of denoising on the reproducibility of radiomic features.

NSCLC Radiogenomics is a radiogenomic dataset from a cohort of 211 patients with non-small cell lung cancer [43], from which we selected the low dose CT images, their respective segmentation masks and clinical data for survival prediction

(n=106). Flowcharts describing the sample selection process from the two datasets and the indices of the included samples

are included in the supplementary material. The average radiation exposure of samples selected from NSCLC

Radiogenomics is 38.65±81.97 mAs (±=standard error of the mean, SEM) (the distribution of radiation exposure for selected samples can be found in Supplementary Figure 2).

*2.4 Experiments*

We trained three cycle GANs to denoise low CT scans: on a paired dataset with low dose CT scans simulated from high dose CT scans with and without the Slice-paired training Strategy strategy (referred to as ablation study hereafter) and on unpaired real low and high dose CT scans.

Then, we assessed the performance of the denoising using Root Mean Square Error (RMSE) and perceptual loss as evaluation metrics. The definition of perceptual loss can be found in equation (3) and definition of RMSE is as follows:

$$RMSE = \sqrt{\frac{1}{M}\sum_{i=1}^{M}(y_i - \hat{y}_i)^2} \tag{5}$$

Where $y_i$ and $\hat{y}_i$ represent the image value in position $i$ for the original high dose CT and denoised CT, respectively. Image values were normalized to 0-1 before calculating RMSE. M represents the number of pixels in one image, 512*512 in our case.

We also assessed the impact of denoising on reproducibility of radiomic features by calculating the concordance correlation coefficients (CCC), as defined in [39]. On the simulated paired data, we calculated the CCCs of the radiomic features extracted in the original high dose CT and the denoised CT. In RIDER, we calculated the CCC of the same day denoised CT scans.

In the ablation study, we assessed the impact of using the position-based training strategy comparing the performance in terms of RMSE, perceptual loss and CCC on synthetic data.

Next, we applied the trained cycle GAN to two applications -- radiomics reproducibility in same-day repeat CT scans and pre-treatment survival prediction – without retraining. Pre-treatment survival prediction of cancer patients is a typical application of radiomics since it appeared in the seminal article by Aerts et al. [2]. We used least squares support vector machines (SVMs) with Radial Basis Function (RBF) Kernel as our classifier. For hyperparameter search and internal validation, we used 40-repeat nested 5-fold cross validation [50]. More details on the survival prediction modelling can be found in [26]. The main metric used for measuring the performance of pre-treatment survival prediction is the area under the receiver operating characteristic curve (AUC) as described in [47].

All experiments were implemented in Python 3.6 and TensorFlow 1.13.1. The training was run on one Nvidia Tesla V100 GPU 30.5GB of memory and 4 CPUs. We set $\lambda$ in equation (4) to 10 and the batch size to 1. The discriminator and the denoiser both used the Adam optimizer [48] and shared the same learning rate. The initial learning rate was set to 0.0002 with a decay factor of 0.8 every 20 epochs. Training runs were stopped at 100 epochs and radiomics features were extracted every 25 epochs (i.e., at 25, 50, 75 and 100 epochs). Table 1 offers a concise summary of our experiments.

Table 1. Summary of Experiment and Corresponding Datasets

| Experiment | Training Strategy | Training Dataset | Testing Dataset |
|---|---|---|---|
| Simulation Data based Training | With Strategy* | Part of paired high-noise and full dose **Lung 1** dataset (n=40, 4260 Frames) | The rest of **high-noise CTs** (n=117, 13423 Frames), **low-noise CTs** (n=157, 17683 Frames) |
| Ablation Study | Without Strategy | Part of paired high-noise and full dose **Lung 1** dataset (n=40, 4260 Frames) | The rest of **high-noise CTs** (n=117, 13423 Frames), **low-noise CTs** (n=157, 17683 Frames) |
| Applications with simulation data training-based networks | With Strategy | Training finished at first part of experiment without re-training | **RIDER** (n=31, 14875 Frames), **NSCLC Radiogenomics** (n=106, 28404 Frames) |
| Applications with real data training-based networks | With Strategy | Low dose CTs from **LIDC-IDRI** (n=12, 3144 Frames), Full dose CTs from **TCGA-LUAD** (n= 14, 3307 Frames) | **RIDER** (n=31, 14875 Frames), **NSCLC Radiogenomics** (n=106, 28404 Frames) |

* means model training with slice-paired training strategy and 'Without Strategy' means model training without slice-paired training strategy.

*2.5 Radiomics Extraction*

The masks of the regions of interest (ROIs) are stored in DICOM format in the Lung 1, RIDER and NSCLC Radiogenomics datasets. The 3D masks for the ROIs are reconstructed from their corresponding files before feature extraction. We used pyradiomics [46] (version 2.2.0) to extract 103 radiomic features for further analysis (full list of features and settings used for pyradiomics can be found in the supplementary Table 5). Shape-related features are not affected by denoising and therefore were excluded from feature reproducibility analysis (removing when testing datasets are low and high noisy CTs and RIDER), resulting in 90 included features. All 103 features were used to derive the 4-year pre-treatment survival prediction model.

# 3. Results

In this section, we present the results of our experiments.

Training the cycle GAN from simulated and real data took 96 and 72 hours respectively. The loss of the generator during training is shown in Figure 2.

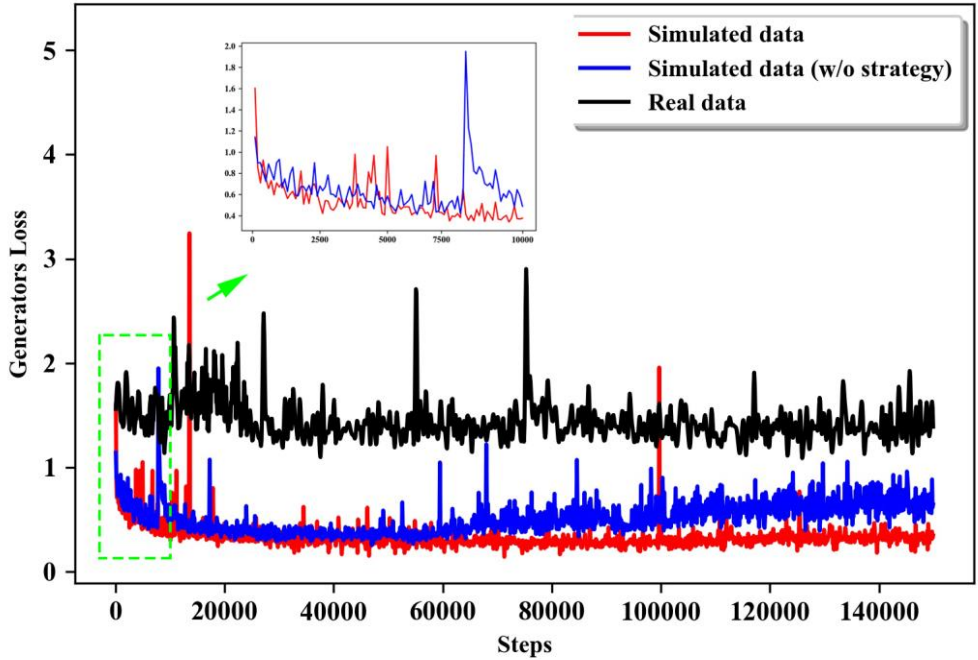

Figure 2. Generator loss over time for different cycle GAN training runs (simulation data trained networks with or without slice-paired training strategy and real data trained network) and zoomed part of generator loss for simulation data trained networks

*3.1 Reproducibility of Radiomic Features on Simulated Paired Data*

An example of an original, noisy and denoised CT scan is shown in Figure 4. We reuse results of CGAN and encoder-decoder network from [13] for better comparison with the cycle GAN (corresponding Figure for high noise image is Supplementary Figure 3). In addition, Table 2 shows the RMSE, perceptual loss and ratio of radiomic features with poor (CCC<0.65), medium ($0.65 \leqslant$ CCC<0.85), and good (CCC $\geqslant 0.85$) reproducibility [49].

As shown in Table 2, the RMSE and perceptual loss of low-noise and high-noise images (before denoising) are 0.0225/0.0706 and 0.0237/0.0781 respectively. Cycle GAN can reduce RMSE and perceptual loss to 0.0170/0.0216 and 0.0181/0.0245 for low-noise and high-noise images. The cycle GAN resulted in higher RMSE than the CGAN but lower perceptual loss, and outperformed the encoder-encoder network in both metrics. The mean CCCs for cycle GAN denoised images improved from 0.87 [95%CI, (0.833,0.901)] and 0.68 [95%CI, (0.617,0.745)] to 0.93 [95%CI, (0.916, 0.949)] and 0.94 [95%CI, (0.928,0.954)] for low-noise images and high-noise images, respectively. A heatmap of radiomics improvement from denoised low-noise images by comparing with original noisy images is shown in Figure 3.

In contrast, encoder-decoder network and CGAN can improve the mean CCC of radiomic features to 0.92 [95%CI, (0.909,0.936)] for low and high-noise images. The cumulative distribution function (CDF) of CCCs for different models when trained for 100 epochs is shown in Figure 4(a-b).

The second investigation of the simulation study was the effect of different training epochs to radiomics reproducibility. The CDF of CCCs for cycle GAN trained at 25, 50, 75 and 100 epochs are shown in Supplementary Figure 4 (a-b). Summary of RMSE, perceptual loss and CCCs of cycle GAN trained at different epochs can be found in Supplementary Table 6. We compared the CCC distributions of radiomic features calculated on images denoised from high-noise images with those of images denoised from low-noise images using the Wilcoxon signed-rank test resulting in a p-value of 0.94. The results show that a cycle GAN trained to denoise high-noise images can be applied to denoise images with different levels of noise and achieve similar results to a CGAN and encoder-decoder network based denoiser[13]. Moreover, we compared the CCC distributions from cycle GAN with CGAN and encoder-decoder network by using the Wilcoxon signed-rank test which resulted in p-values of 0.73 and 0.07, respectively. The results show that a cycle GAN achieved similar results to CGAN and encoder-decoder networks, and that in some cases, Cycle Gan even received better results.

Figure 3. A heatmap of radiomics improvement from denoised low-noise images, results about EDN and CGAN

reproduced from [13]

Table 2. Summary of RMSE, perceptual loss and distribution of CCCs of radiomic features based on denoising simulated datasets.

| Models \ Distribution | RMSE | Perceptual loss | CCCs<0.65 | 0.65≤CCCs<0.85 | CCCs≥0.85 |
|---|---|---|---|---|---|
| Low-noise Images | | | | | |
| Without denoising | 0.0225 | 0.0706 | 10% | 22% | 68% |
| Encoder-decoder | 0.0173 | 0.0427 | 0% | 19% | 81% |
| CGAN | 0.0143 | 0.0290 | 3% | 17% | 80% |
| Cycle GAN | 0.0170 | 0.0216 | 0% | 16% | 84% |
| Cycle GAN (w/o strategy) | 0.0167 | 0.0258 | 1% | 13% | 86% |
| High-noise Images | | | | | |
| Without denoising | 0.0237 | 0.0781 | 36% | 23% | 41% |
| Encoder-decoder | 0.0175 | 0.0443 | 4% | 16% | 80% |
| CGAN | 0.0146 | 0.0305 | 0% | 16% | 84% |
| Cycle GAN | 0.0181 | 0.0245 | 0% | 14% | 86% |
| Cycle GAN (w/o strategy) | 0.0188 | 0.0256 | 3% | 12% | 84% |

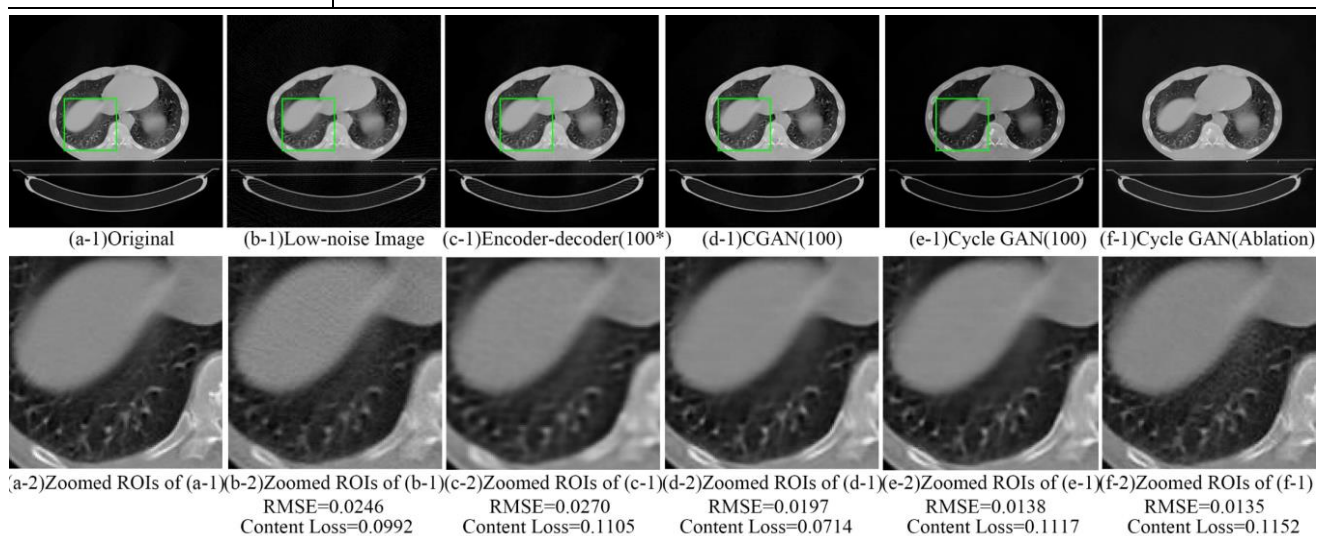

(a-1)Original   (b-1)Low-noise Image   (c-1)Encoder-decoder(100*)   (d-1)CGAN(100)   (e-1)Cycle GAN(100)   (f-1)Cycle GAN(Ablation)

(a-2)Zoomed ROIs of (a-1)   (b-2)Zoomed ROIs of (b-1)   (c-2)Zoomed ROIs of (c-1)   (d-2)Zoomed ROIs of (d-1)   (e-2)Zoomed ROIs of (e-1)   (f-2)Zoomed ROIs of (f-1)

|  | RMSE=0.0246 Content Loss=0.0992 | RMSE=0.0270 Content Loss=0.1105 | RMSE=0.0197 Content Loss=0.0714 | RMSE=0.0138 Content Loss=0.1117 | RMSE=0.0135 Content Loss=0.1152 |

Figure 4. Example of low dose CT denoising. (a-1) The original full dose CT image; (b-1) Low-noise image; (c-1) Image denoised by encoder-decoder network (*Training at 100 epochs); (d-1) Image denoised by CGAN; (e-1) Image denoised by cycle GAN; (f-1) Image denoised by cycle GAN (ablation study); (a-2) to (f-2) Zoomed ROIs for (a-1) to (f-1).

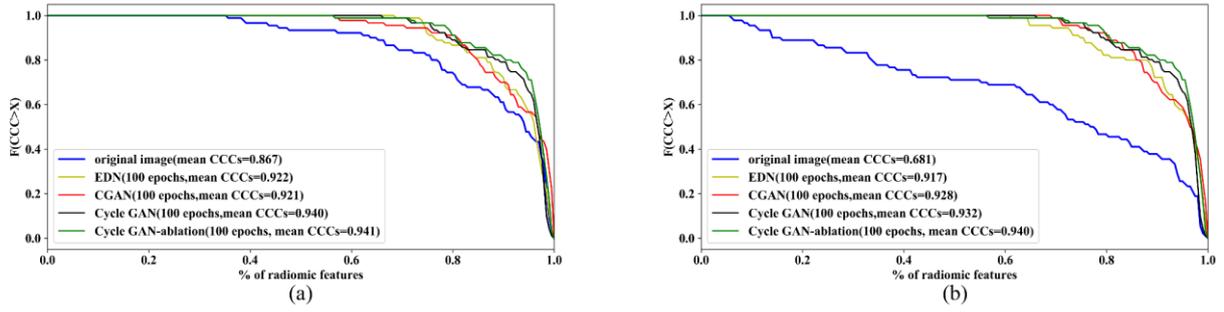

Figure 5. CDF of CCC by Using cycle GAN. (a) CDF of CCC based on denoised low-noise images by using different

models; (b) CDF of CCC based on denoised high-noise images by using different models.

*3.2 Ablation Study for the Training Strategy*

An example of denoised images from cycle GAN ablation study can be found in Figure 4 (f-1) and Figure 4 (f-2).

Table 2 and Supplementary Table 7 shows the RMSE, perceptual loss and ratio of poor, medium, and good reproducibility radiomic features about ablation study of cycle GAN. The cycle GAN trained without our training strategy can also reduce the RMSE and perceptual loss of low-noise and high-noise images to 0.0167/0.0258 and 0.0188/0.0256 respectively. Moreover, it can increase the average CCC to 0.94 [95%CI, (0.924,0.957)] and 0.93 [95%CI, (0.917,0.953)]

for low and high-noise images respectively. The CDF of CCCs for ablation study when trained for 100 epochs is shown in from ablation study trained at 100 epochs was compared with results from a network trained with training strategy and we found no signification differences (Wilcoxon signed-rank test, p-value=0.11). Figure **2** shows that training the cycle GAN with the training strategy might speed up convergence slightly. On the other hand, without the training strategy, the generator's loss function increases beyond 60000 steps. Finally, the cycle GAN trained with our training strategy led to significantly higher CCCs when trained for only 25 epochs (Wilcoxon signed-rank test, p-value < 0.01), as shown comparing Supplementary Figure 4(a) to (c) and Figure 4(b) to 4(d).

Figure **5** (a-b) and the differences among epochs can be found in Supplementary Figure 4 (c-d). The distribution of CCCs

Our results seems different from research reported elsewhere [41] which found that slide-based training strategy can improve denoising performance. We will discuss this point more in the discussion section.

*3.3 Reproducibility on Real Data*

We now focus on the impact of denoising on the reproducibility of radiomic features in same day repeat low dose CT scans (RIDER dataset). An example of an original image and its denoised counterparts denoised using a CGAN, an encoder-decoder network and the cycle GANs trained on simulated and real data are shown in Figure 6. Figure 7 shows the CDF of the CCCs for the radiomic features extracted from the original and denoised CT images. The cycle GAN trained on real data outperforms the rest of generative models (Wilcoxon signed-rank test, p-value < 0.01). On the other hand, the performance of the cycle GAN trained on simulated data is similar to that of the encoder-decoder network and CGAN ( p-value = 0.23 and 0.56 for respectively).

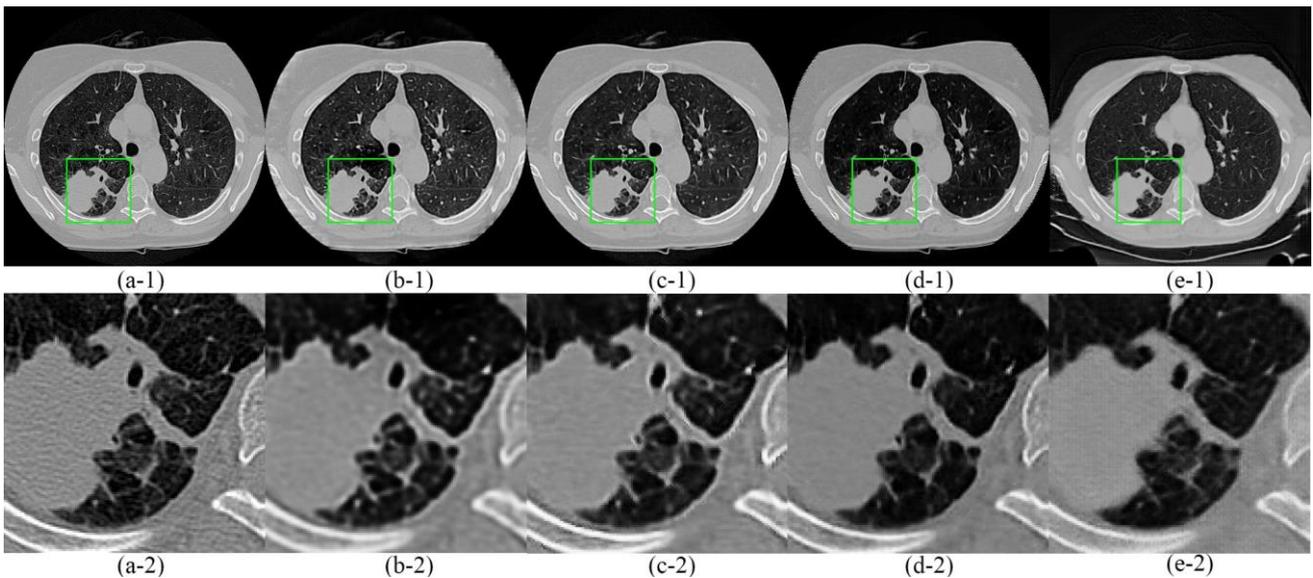

Figure 6. Example of RIDER denoising. (a-1) One original image from RIDER; (b-1) Image denoised by encoder-decoder network (Training at 100 epochs); (c-1) Image denoised by CGAN (Training at 100 epochs); (d-1) Image denoised by simulation data trained cycle GAN (Training at 100 epochs); (e-1) Image denoised by real data trained cycle GAN (Training at 100 epochs); (a-2) to (e-2) Zoomed ROIs for (a-1) to (e-1).

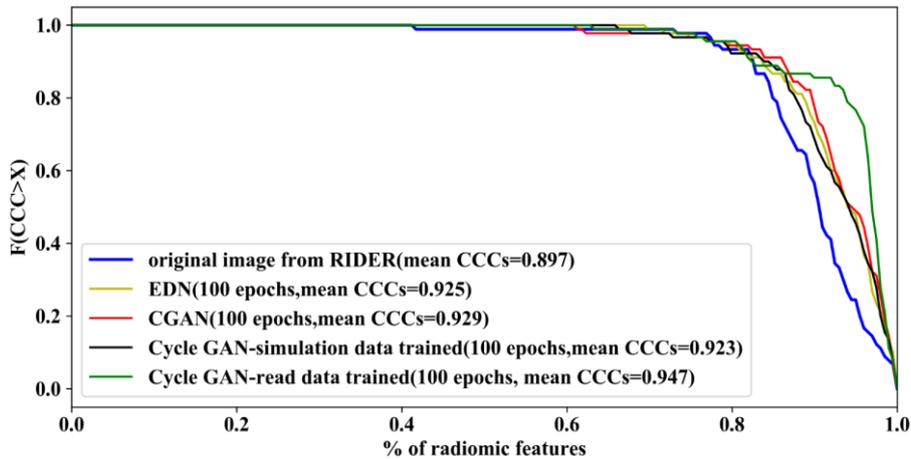

Figure 7. CDF of CCCs and for denoised CT scans in the RIDER dataset.

*3.4 Survival prediction on Real Data*

An example of an original NSCLC Radiogenomics image, and its denoised counterparts based on CGAN, encoder-decoder network and cycle GANs trained from simulated and real data can be found in Supplementary Figure 5.

Figure 8 illustrates the results of the of 4-year pre-treatment survival prediction experiment showing the AUC for each generative model across different number of epochs. We achieved an AUC for survival prediction based on radiomics extracted from the original NSCLC Radiogenomics dataset of 0.52 [95%CI, (0.511,0.538)] at 100 epochs. Denoising the CT scans using a CGAN or an encoder-decoder network led to models with an increased AUC of 0.57 [95%CI, (0.551, 0.580)] (at 100 epochs) as shown in [26]. The cycle GANs trained on simulated and real data resulted in a higher mean AUC of around 0.58 [95%CI, (0.576,0.596)] but the difference between models was not statistically significant (Student's t-test, all p-values > 0.10).

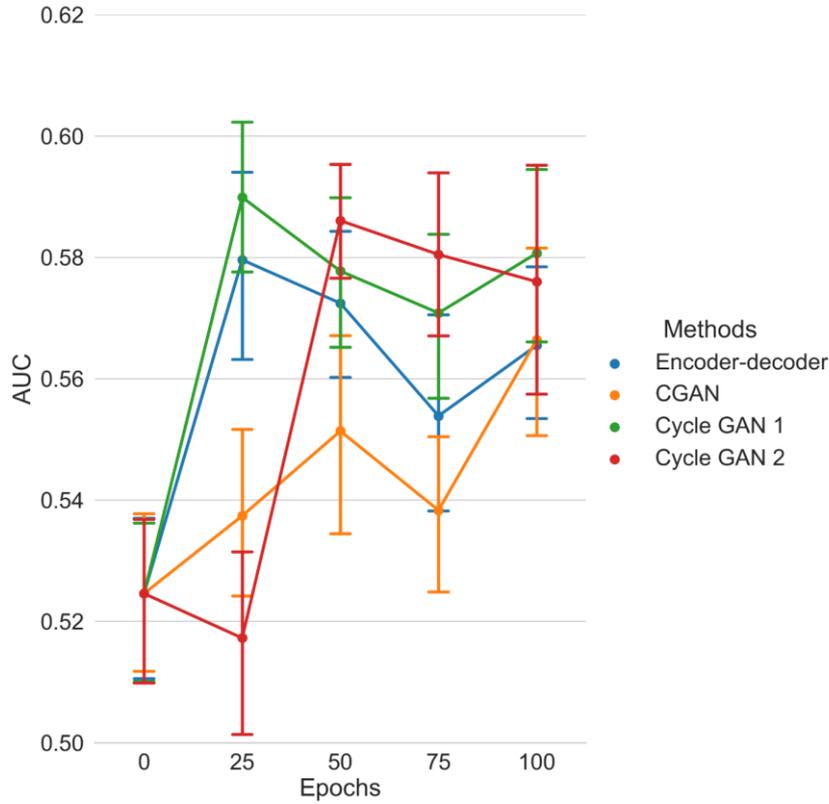

Figure 8. Results (AUC) of 4-year pre-treatment survival prediction.

Cycle GAN 1 and 2 were trained on simulated and real data, respectively. To interpret the improvement of AUC in 4-year survival prediction tasks, we used an RBF kernel based SVM Recursive Feature Elimination algorithm [53] to assess the importance of features in the prediction model. Table 3 shows the top eight most important features in the models trained on the radiomic features from the original images and those from denoised images (The table with all features can be found in Supplementary Table 7). Six features appeared in all four models (highlighted in green in Figure 3). These features' CCC improved by denoisers, most of them improved significantly, which might explain how denoising can improve the AUC of survival prediction models.

Table 3. Top eight most important features in the survival prediction model trained on noisy images and images denoised using different generative models

| Rank | Original images | Denoised with EDN | Denoised with CGAN | Denoised with Cycle GAN |
|------|-----------------|-------------------|--------------------|-------------------------|
| 1 | glszm_LargeArea | glszm_LargeArea | glszm_LargeArea | glrlm_GrayLevel |

|   | | | |
|---|---|---|---|
|   | LowGrayLevelEmphasis | LowGrayLevelEmphasis | LowGrayLevelEmphasis | NonUniformityNormalized |
| 2 | ngtdm_Coarseness | gldm_GrayLevelVariance | glrlm_GrayLevel NonUniformityNormalized | glszm_LargeArea LowGrayLevelEmphasis |
| 3 | gldm_GrayLevelVariance | glszm_LargeArea LowGrayLevelEmphasis | gldm_GrayLevelVariance | gldm_GrayLevelVariance |
| 4 | firstorder_Energy | gldm_LargeDependence HighGrayLevelEmphasis | firstorder_Energy | firstorder_Energy |
| 5 | shape_MinorAxisLength | gldm_GrayLevel NonUniformity | gldm_GrayLevel NonUniformity | shape_MinorAxisLength |
| 6 | glrlm_GrayLevel NonUniformityNormalized | firstorder_Energy | ngtdm_Coarseness | ngtdm_Coarseness |
| 7 | glszm_LargeArea HighGrayLevelEmphasis | glcm_JointEntropy | glcm_JointEntropy | glcm_JointEntropy |
| 8 | glcm_JointEntropy | ngtdm_Coarseness | shape_MinorAxisLength | glrlm_RunLength NonUniformityNormalized |

## 4. Discussion

The objective of our study was to investigate the potential of cycle GANs for denoising low dose CTs to improve the reproducibility of radiomics features and the performance of radiomics-based models. For this purpose, we trained two cycle GANs, one with simulated paired data and the other one with real data, to denoise low dose CT scans. In order to measure the performance of our denoising models, we ran experiments and compared the results of our method with those of CGANs and encoder-decoder networks trained on simulated paired data. The results show that both cycle GANs trained on simulated and on real data can improve radiomics' reproducibility and performance in low dose CT and achieve similar results compared to CGANs and encoder-decoder networks.

The main advantage of cycle GANs over CGANs and encoder-decoder networks is that they do not required paired images, which are virtually impossible to collect. For CGANs and encoder-decoder networks we overcame this issue by generating simulated low dose CTs by introducing noise into high dose CTs [Simulation study]. However, simulated noise might differ from noise encountered in low dose CTs. Hence, being able to train a model on real low dose CT scans is a significant advantage. However, training cycle GANs is volatile, especially when the target domain and the source domain differ, as documented elsewhere [27][51]. Ideally, in order to maximize the chances of success for the training process, training data would be collected from the same scanner, with the same protocol (except radiation exposure), and from the same group of patients for the two domains (low and high dose CT). However, such a dataset is not available to us. Hence, we defined selection criteria for the training data so that the source and target image domains kept certain similarities. We chose scanner manufacturer and table height (which determines field of view and the height of human body) based on [12]. These inclusion criteria were introduced after several failed attempts at training a cycle GAN with the full dataset. Examples of failed training runs are shown in Figure 9. However, trained models retain certain generalizability and can achieve good results across different scanners with different parameter settings as shown in the results (images in the RIDER and NSCLC Radiogenomics datasets were scanned from multiple types of scanners with different protocols).

The slice-paired training strategy we proposed seems to lead to slightly faster convergence as hinted by the loss plot and the models' results at 25 epochs. However, this strategy did not lead to significant improvement of the networks' denoising performance at 100 epochs. One possible explanation is that the training strategy cannot make the resulting network a better approximator of the mapping from low dose CT domain to high dose Figure 2 and the comparisons between Supplementary Figure 4 (a) to (c) and (b) to (d) seem to support this view. Another possible hypothesis for this phenomenon is that reproducibility and performance of radiomics may not be so sensitive to the quality of images when the quality reaches a certain threshold.

As mentioned above, cycle GANs achieved similar performance to CGAN and encoder-decoder network trained on simulated data, slightly outperforming them in some experiments. The difference in performance might be explained by the differences in the architectures used: the generator in CGAN and the encoder-decoder is a 5-layer network while there are 9 ResNet blocks [52] (27 convolutional layers) in the cycle GAN's generators. Related articles have hypothesized [22] that neural networks for 'low level' domain adaptation – such as denoising – should be kept shallow, since texture transfer in 'low level' domain adaptation is not significant. However, the results in our study seem to show that very deep neural network can also achieve good performance in some 'low level' domain adaptation tasks.

Our study suffered from a few limitations. First, there were important differences between the populations in different training datasets (LIDC-IDRI and TCGA-LUAD). For example, patients in TCGA-LUAD were thinner than patients in LIDC-IDRI, as shown in Supplementary Figure 4. Hence, the cycle GAN trained on these datasets learnt to not only denoise the images, but make the patients thinner as illustrated in Figure **6** (e-1). Fortunately, the ROIs of this study are located in the lung and the volume of patients' lung in two domains are similar. Therefore, there was no significant size shift in the ROIs. Second, due to the differences of the CT bed in LIDC-IDRI and TCGA-LUAD, the cycle GAN also transforms bottom part of the image as shown in Figure **6** (e-1). Third, the cycle GAN trained on real data performed poorly on simulated noisy images in terms of improving the reproducibility of radiomic features. However, we believe that the good performance in real data is more important than the performance in simulated data, since it is more representative of real applications. Fourth, one of the deductions of our slice-paired training strategy ' the first slice of a low dose CT scan will have higher similarity with the first slice of a high dose CT scan' is not automatically true. The similarity of first slice of a CT scan depends on a lot of factors such as the patient position, section of the body is scanned etc., these factors were ignored in this paper.

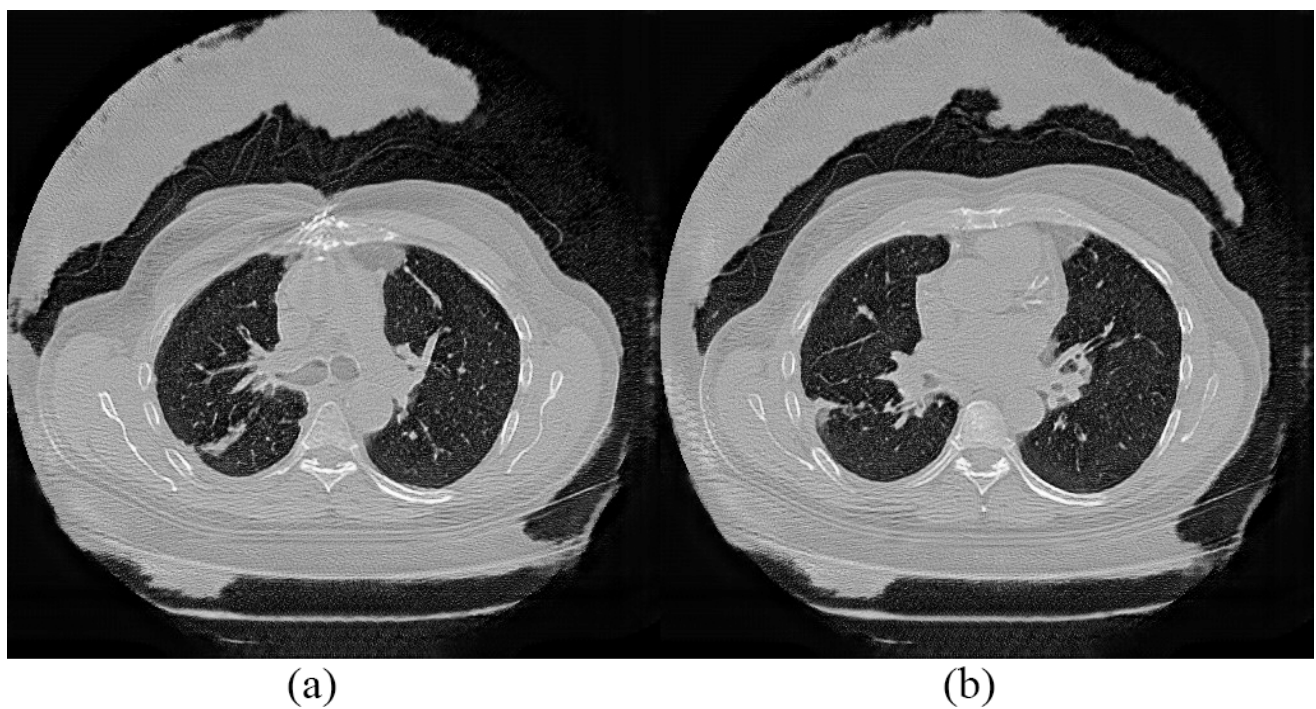

(a)                                                          (b)

Figure 9. Examples of failed cycle GAN training.

## 5. Conclusions

In this study, we investigate the potential of denoising low dose CT using cycle GANs to improve the reproducibility of radiomics features and the performance of radiomics based prediction models. We trained two cycle GANs: using paired simulated low dose CTs and unpaired real low and high CT images. To accelerate convergence, we introduced a slice-paired training strategy.

The results of our experiments show that a cycle GAN trained to denoise low dose CT scans from unpaired low and high dose CT scans can improve the reproducibility of radiomic features in simulated low dose CTs and same-day repeat low dose CTs. In addition, we showed that radiomics based pre-treatment survival prediction models trained on low dose CT scans denoised with said cycle GAN can achieve better performance. The improvement in reproducibility and prediction model performance are comparable to those achieved with CGANs and encoder decoder networks trained on simulated paired data. Cycle GANs have better potential because they do not need paired data, but they are burdened by the volatility of the treatment process, which limits their applicability. More research is needed to make cycle GAN training more robust, for them to be able to be trained on a more diverse dataset.

## 6. Acknowledgements

J Chen is supported by a China Scholarship Council scholarship (201906540036). The remaining authors acknowledge funding support from the following: STRaTegy (STW 14930), BIONIC (NWO 629.002.205), TRAIN (NWO 629.002.212), CARRIER (NWO 628.011.212) and a personal research grant by The Hanarth Funds Foundation for L Wee.

## 7. Conflict of Interest

The authors declare no conflicts of interest.

## 8. References

[1] La Thangue, Nicholas B., and David J. Kerr. "Predictive biomarkers: a paradigm shift towards personalized cancer medicine." *Nature reviews Clinical oncology* 8.10 (2011): 587-596.

[2] Aerts, Hugo JWL, et al. "Decoding tumour phenotype by noninvasive imaging using a quantitative radiomics approach." *Nature communications* 5.1 (2014): 1-9.

[3] Desseroit, Marie-Charlotte, et al. "Reliability of PET/CT shape and heterogeneity features in functional and morphologic components of non–small cell lung cancer tumors: a repeatability analysis in a prospective multicenter cohort." *Journal of Nuclear Medicine* 58.3 (2017): 406-411.

[4] Bogowicz, Marta, et al. "Stability of radiomic features in CT perfusion maps." *Physics in Medicine & Biology* 61.24 (2016): 8736.

[5] Tixier, Florent, et al. "Reproducibility of tumor uptake heterogeneity characterization through textural feature analysis in 18F-FDG PET." *Journal of Nuclear Medicine* 53.5 (2012): 693-700.

[6] Li, Hui, et al. "Quantitative MRI radiomics in the prediction of molecular classifications of breast cancer subtypes in

the TCGA/TCIA data set." *NPJ breast cancer* 2.1 (2016): 1-10.

[7] Leandrou, Stephanos, et al. "Quantitative MRI brain studies in mild cognitive impairment and Alzheimer's disease: a methodological review." *IEEE reviews in biomedical engineering* 11 (2018): 97-111.

[8] Chaddad, Ahmad, Christian Desrosiers, and Matthew Toews. "Multi-scale radiomic analysis of sub-cortical regions in MRI related to autism, gender and age." *Scientific reports* 7.1 (2017): 1-17.

[9] Bodalal, Zuhir, et al. "Radiogenomics: bridging imaging and genomics." *Abdominal radiology* 44.6 (2019): 1960-1984.

[10] Berenguer, Roberto, et al. "Radiomics of CT features may be nonreproducible and redundant: influence of CT acquisition parameters." *Radiology* 288.2 (2018): 407-415.

[11] Welch, Mattea L., et al. "Vulnerabilities of radiomic signature development: the need for safeguards." *Radiotherapy and Oncology* 130 (2019): 2-9.

[12] Meyer, Mathias, et al. "Reproducibility of CT radiomic features within the same patient: influence of radiation dose and CT reconstruction settings." *Radiology* 293.3 (2019): 583-591.

[13] Chen, Junhua et al. "Generative Models Improve Radiomics Reproducibility in Low Dose Cts: A Simulation Study." *Physics in Medicine & Biology* 66.16 (2021): 165002.

[14] Sagheer, Sameera V. Mohd, and Sudhish N. George. "A review on medical image denoising algorithms." *Biomedical signal processing and control* 61 (2020): 102036.

[15] Rudin, Leonid I., Stanley Osher, and Emad Fatemi. "Nonlinear total variation based noise removal algorithms." *Physica D: nonlinear phenomena* 60.1-4 (1992): 259-268.

[16] Chan, Raymond H., Chung-Wa Ho, and Mila Nikolova. "Salt-and-pepper noise removal by median-type noise detectors and detail-preserving regularization." *IEEE Transactions on image processing* 14.10 (2005): 1479-1485.

[17] Chen, Yunjin, and Thomas Pock. "Trainable nonlinear reaction diffusion: A flexible framework for fast and effective image restoration." *IEEE transactions on pattern analysis and machine intelligence* 39.6 (2016): 1256-1272.

[18] Zhang, Kai, et al. "Beyond a gaussian denoiser: Residual learning of deep cnn for image denoising." *IEEE transactions on image processing* 26.7 (2017): 3142-3155.

[19] Zhang, Kai, et al. "Learning deep CNN denoiser prior for image restoration." *Proceedings of the IEEE conference on computer vision and pattern recognition*. 2017.

[20] Krizhevsky, Alex, Ilya Sutskever, and Geoffrey E. Hinton. "Imagenet classification with deep convolutional neural networks." *Advances in neural information processing systems* 25 (2012): 1097-1105.

[21] Lefkimmiatis, Stamatios. "Non-local color image denoising with convolutional neural networks." *Proceedings of the IEEE Conference on Computer Vision and Pattern Recognition*. 2017.

[22] Chen, Hu, et al. "Low-dose CT with a residual encoder-decoder convolutional neural network." *IEEE transactions on medical imaging* 36.12 (2017): 2524-2535.

[23] Yang, Qingsong, et al. "Low-dose CT image denoising using a generative adversarial network with Wasserstein distance and perceptual loss." *IEEE transactions on medical imaging* 37.6 (2018): 1348-1357.

[24] Goyal, Bhawna, et al. "Image denoising review: From classical to state-of-the-art approaches." *Information fusion* 55 (2020): 220-244.

[25] Li, Yajun, et al. "Normalization of multicenter CT radiomics by a generative adversarial network method." *Physics in Medicine & Biology* 66.5 (2021): 055030.

[26] Chen, Junhua, et al. " Generative Models Improve Radiomics Performance in Different Tasks and Different Datasets: An Experimental Study" *arXiv preprint arXiv: 2109.02252* (2021).

[27] Zhu, Jun-Yan, et al. "Unpaired image-to-image translation using cycle-consistent adversarial networks." *Proceedings*

*of the IEEE international conference on computer vision*. 2017.

[28] Yang, Wei, et al. "Improving low-dose CT image using residual convolutional network." *IEEE access* 5 (2017): 24698-24705.

[29] Chen, Hu, et al. "Low-dose CT with a residual encoder-decoder convolutional neural network." *IEEE transactions on medical imaging* 36.12 (2017): 2524-2535.

[30] Yang, Qingsong, et al. "Low-dose CT image denoising using a generative adversarial network with Wasserstein distance and perceptual loss." *IEEE transactions on medical imaging* 37.6 (2018): 1348-1357.

[31] Mirza, Mehdi, and Simon Osindero. "Conditional generative adversarial nets." *arXiv preprint arXiv:1411.1784* (2014).

[32] Manning, Christopher, and Hinrich Schutze. *Foundations of statistical natural language processing*. MIT press, 1999.

[33] Olkin, Ingram, and Friedrich Pukelsheim. "The distance between two random vectors with given dispersion matrices." *Linear Algebra and its Applications* 48 (1982): 257-263.

[34] Johnson, Justin, Alexandre Alahi, and Li Fei-Fei. "Perceptual losses for real-time style transfer and super-resolution." *European conference on computer vision*. Springer, Cham, 2016.

[35] McCollough, C.H., et al. (2020). Low Dose CT Image and Projection Data [Data set]. *The Cancer Imaging Archive*. https://doi.org/10.7937/9npb-2637.

[36] Zhao, Binsheng, et al. "Evaluating variability in tumor measurements from same-day repeat CT scans of patients with non–small cell lung cancer." *Radiology* 252.1 (2009): 263-272.

[37] Kang, Eunhee, et al. "Cycle-consistent adversarial denoising network for multiphase coronary CT angiography." *Medical physics* 46.2 (2019): 550-562.

[38] Isola, Phillip, et al. "Image-to-image translation with conditional adversarial networks." *Proceedings of the IEEE*

*conference on computer vision and pattern recognition*. 2017.

[39] Lawrence, I., and Kuei Lin. "A concordance correlation coefficient to evaluate reproducibility." *Biometrics* (1989): 255-268.

[40] Chen, Junhua, et al. "Lung Cancer Diagnosis Using Deep Attention Based on Multiple Instance Learning and Radiomics." *arXiv preprint arXiv:2104.14655* (2021).

[41] Yang, Heran, et al. "Unsupervised MR-to-CT Synthesis Using Structure-Constrained CycleGAN." *IEEE transactions on medical imaging* 39.12 (2020): 4249-4261.

[42] Aerts, H. J. W. L., Wee, L., Rios Velazquez, E., Leijenaar, R. T. H., Parmar, C., Grossmann, P., … Lambin, P. (2019). Data From NSCLC-Radiomics [Data set]. The Cancer Imaging Archive. https://doi.org/10.7937/K9/TCIA.2015.PF0M9REI

[43] Bakr, Shaimaa, et al. "A radiogenomic dataset of non-small cell lung cancer." *Scientific data* 5.1 (2018): 1-9.

[44] Armato III, Samuel G., et al. "The lung image database consortium (LIDC) and image database resource initiative (IDRI): a completed reference database of lung nodules on CT scans." *Medical physics* 38.2 (2011): 915-931.

[45] Albertina, B., Watson, M., Holback, C., Jarosz, R., Kirk, S., Lee, Y., … Lemmerman, J. (2016). Radiology Data from The Cancer Genome Atlas Lung Adenocarcinoma [TCGA-LUAD] collection. The Cancer Imaging Archive. http://doi.org/10.7937/K9/TCIA.2016.JGNIHEP5

[46] Van Griethuysen, Joost JM, et al. "Computational radiomics system to decode the radiographic phenotype." *Cancer research* 77.21 (2017): e104-e107.

[47] Huang, Jin, and Charles X. Ling. "Using AUC and accuracy in evaluating learning algorithms." *IEEE Transactions on knowledge and Data Engineering* 17.3 (2005): 299-310.

[48] Kingma, Diederik P., and Jimmy Ba. "Adam: A method for stochastic optimization." *arXiv preprint*

*arXiv:1412.6980* (2014).

[49] Traverso, Alberto, et al. "Repeatability and reproducibility of radiomic features: a systematic review." *International Journal of Radiation Oncology\* Biology\* Physics* 102.4 (2018): 1143-1158.

[50] Cawley, Gavin C., and Nicola LC Talbot. "On over-fitting in model selection and subsequent selection bias in performance evaluation." *The Journal of Machine Learning Research* 11 (2010): 2079-2107.

[51] J. Lee, J. Gu and J. C. Ye, "Unsupervised CT Metal Artifact Learning using Attention-guided β-CycleGAN," in *IEEE Transactions on Medical Imaging*, 2021 doi: 10.1109/TMI.2021.3101363.

[52] He, Kaiming, et al. "Deep residual learning for image recognition." *Proceedings of the IEEE conference on computer vision and pattern recognition*. 2016.

[53] Liu, Quanzhong, et al. "Feature selection for support vector machines with RBF kernel." *Artificial Intelligence Review* 36.2 (2011): 99-115.